\documentclass[prb,twocolumn,showpacs,preprintnumbers,amsmath,amssymb]{revtex4}

\usepackage{graphicx}
\usepackage{dcolumn}
\usepackage{bm}

\begin{document}

\title{Many-body approach to infinite non-periodic systems:\\
application to the surface of semi-infinite jellium}

\author{G. Fratesi}
\altaffiliation{Present address: SISSA/ISAS, via Beirut 4, I-34014
Trieste} \email{fratesi@sissa.it}
\author{G. P. Brivio} \email{gianpaolo.brivio@mater.unimib.it}
\affiliation{INFM and Dipartimento di Scienza dei Materiali,\\
Universit\`a di Milano-Bicocca, via Cozzi 53, I-20125 Milano}
\author{L. G. Molinari} \affiliation{Dipartimento di Fisica
dell'Universit\`a degli Studi di Milano\\ Via Celoria 16, I-20133
Milano} \email{luca.molinari@mi.infn.it}

\date{\today}

\begin{abstract}
A method to implement the many-body Green function formalism in the
$GW$ approximation for infinite non periodic systems is presented.
It is suitable to treat systems of known ``asymptotic'' properties
which enter as boundary conditions, while the effects of the lower
symmetry are restricted to regions of finite volume.
For example, it can be applied to surfaces or localized impurities.
We illustrate the method with a study of the surface of semi-infinite
jellium.
We report the dielectric function, the effective potential and the
electronic self-energy discussing the effects produced by the screening
and by the charge density profile near the surface.
\end{abstract}


\maketitle

\section{\label{sec:introduction}Introduction}

Density functional theory (DFT) is central to the understanding of
several fundamental problems in condensed matter physics, since it
allows for obtaining the ground state properties of the system in a
very accurate and computationally feasible way within an {\it ab
initio} framework.
However, current approximations of the exchange correlation functional
are based mainly on the Local Density Approximation (LDA), and hence
on the properties of the homogeneous electron gas.
Therefore they cannot account correctly for fundamental
physical phenomena, like weak bonding, and consequently the van der
Waals interaction.
At surfaces, the LDA and derived Generalized Gradient
Approximation (GGA) functionals are unsatisfactory and corrections for
surface specific errors were recently proposed.\cite{mat01}
Another example of the failure of the LDA/GGA is the
potential calculated at large distances from a metal surface, which
instead of reproducing the classical image one,\cite{egu92}
displays an exponential decay.
This because the LDA/GGA neglects the long range electron
correlations due to the inhomogeneity of the surface charge density.

A way to improve on the limitations of the DFT solution caused by
approximated exchange-correlation functionals is to resort to methods
of many-body perturbation theory.
This also gives access to physical properties outside the realm of
DFT, such as quasiparticle and collective excitations.\par

The many-body problem was expressed by Hedin\cite{hed65} as a formally
exact closed set of five equations that relate
the single particle Green's function, the self-energy, the
polarization, the effective two-particle potential and the vertex
function.
The $GW$ approximation\cite{aul00,ary98} (GWA) neglects vertex
corrections and reduces Hedin's system to four closed equations.
The results of DFT can be used to achieve an efficient scheme of
solution of the $GW$ equations.
In fact Dyson's equation for the exact Green function requires
the input of a reference one, which in the original Hedin's
formulation is the Green function for the Hartree equation.
A great improvement is to use instead the Green function of the Kohn-Sham
(KS) equation, since it includes not only the Hartree potential but also
exchange and correlation effects to some local degree of correctness.
This minimizes the effort for self-consistency in the evaluation of
the density.

The GWA with the just mentioned input from DFT is a strategy that has
been successfully applied to several systems.
It combines the efficiency of the KS equation, with the use of the
basic equations of many body theory, and it is able to supply
reliable ground state properties as well as excitation energies, lifetimes,
and generalized dielectric functions.
For example the surface effective potential $V_{\textrm{XC}}(z)$ with
the correct image tail was achieved by Eguiluz {\em et
al}.,\cite{egu92} by solving the Sham-Schl{\"u}ter equation in GWA,
with a slab geometry.

As already pointed out, DFT, sometimes coupled to the GWA especially
for excited state properties,\cite{oni02} have allowed for a realistic
and accurate treatment of more and more complex systems of condensed
matter physics.
Among them, a single bulk impurity, solid surfaces, adsorbed molecules and
clusters, interfaces as well as the recently investigated nano-contacts
are indeed infinite or semi-infinite non-periodic electronic systems.
However, in the theoretical description, such systems are usually confined
in a ficticiously finite region/supercell with periodic properties at the
boundaries.\cite{gpb99}
Consequently artificial features may occur: for example, spurious
interactions, non-physical oscillatory behaviors of the wave-functions, and a
discrete spectrum in which it may be difficult to isolate localized
electronic states and resolve resonant ones.
For example, in order to properly account for the underlying bulk band
structure, a DFT evaluation of the surface dielectric function for a
semi-infinite crystal, that avoids slab or supercell geometries, was
recently proposed by Brodersen and Schattke.\cite{bro02}

In this paper we address the problem of the description of an infinite
non-periodic system within a GWA framework.
We show here that the artificial reduction of an infinite volume to a
periodic one can be avoided with minimal computational cost and more
transparent results.
Our method applies to systems (like surfaces or localized impurities)
that asymptotically in space identify with others whose relevant
correlators are well studied, like bulk crystal or vacuum.
It relies on the assumption that, for any relevant correlator $F(1,2)$,
there is a {\em finite} space region $U_F$ outside which $F(1,2)$
cannot be distinguished from its asymptotic limit $F^{\infty}(1,2)$,
within the required accuracy.
This property is under control in the course of computation. It also provides
a powerful tool for checking the correctness of the result, by verifying
whether the correlator $F$ matches with $F^{\infty}$ at the boundaries of
$U_F$ or not.

The virtue of this method is to {\em {replace the finite volume of
the cell or slab by a finite effective volume $U_F$ with boundary
conditions that are intrinsic to the system}}.
The property that ``asymptotic'' regions actually determine an
effective region $U_F$ of finite volume for the computation is
connected with the principle of {\em nearsightedness} recently
introduced by Kohn.\cite{koh96,koh98}
It states that the local electronic structure near a point
$\mathbf{r}$, while requiring in principle the knowledge of the
density (or the effective potential) everywhere, is largely determined
by the potential near $\mathbf{r}$.

In Sec.~\ref{sec:gwa} we shall present the main points of the GWA to
set the stage for further developments.
Sec.~\ref{sec:method} outlines the method.
Two main ingredients are the embedding approach for the zero-th order
Green function, which guarantees that the properties of the infinite
non-periodic system are taken correctly into account, and a Lemma for
the inversion of infinite matrices.
The application to the semi-infinite jellium is worked out in
Sec.~\ref{sec:semiinfj}.
Because of its generality, the jellium surface is
currently used as a bench mark system to evaluate many-body
features.\cite{god02}
The extension of the GWA to the semi-infinite jellium can provide
further data especially on how many-body properties affect the spectral
ones by calculating the dielectric function, the effective potential and
the self-energy for a true continuum.
Finally Sec.~\ref{sec:concl} is devoted to the conclusions.

\section{\label{sec:gwa}The $GW$ approximation}

We wish to carry out a many-body treatment of an infinite non periodic
system in the GWA.
In order to present our work in a self-contained way, in this section we
recall the basic properties and the equations of the GWA.
To be specific, we write the Hamiltonian of a system of electrons with
Coulomb interaction $v({\mathbf{r}},{\mathbf{r}}^\prime)$, in a static
external potential $V_{\textrm{ext}}({\mathbf{r}})$ that couples to
the density $\hat{n}({\mathbf{r}})$:
\begin{eqnarray}
    {\hat H}=\sum_i \frac{1}{2} {\hat{\mathbf p}}_i^2 + \sum_{i<j}
    v(\hat{\mathbf{r}}_i, \hat{\mathbf{r}}_j)+ \int \mathrm{d}\mathbf{r}
    V_{\textrm {ext}}(\mathbf{r}){\hat n}(\mathbf{r}).
\end{eqnarray}
Atomic units ($a_0=0.529$~\AA, 1~hartree$=27.2$~eV) are used
throughout this paper.
Since we are not interested in a spin-polarized phase, we consider a
ground state with equal occupations for spin.
The fermionic correlators are then proportional to the unit spin
matrix.
Two-point correlators, $F(1,2) \equiv
F(\mathbf{r}_1t_1,\mathbf{r}_2t_2)$ are time translation invariant and
will be considered in frequency space:
\begin{eqnarray}\label{eq:time_ft}
    F(\mathbf{r}_1t_1,\mathbf{r}_2t_2) = \int_{-\infty}^{+\infty}
    {\frac{\mathrm{d}\omega}{2\pi}} F(\mathbf{r}_1,\mathbf{r}_2,\omega)
    e^{-i\omega(t_1-t_2)}.
\end{eqnarray}

The GWA\cite{ary98,aul00} is a self-consistent scheme that originates
from truncating the exact closed set of five Hedin's equations for the
five basic quantities: Green function $G$, self-energy $\Sigma $,
effective potential $W$, polarization $P$, and vertex function
$\Gamma$.
In the GWA the first two Hedin's equations, namely the Dyson
equations for the Green function $G$ and for the the effective
potential $W$, remain the same.

The computational effort is reduced if, in the Dyson equation for $G$,
one makes reference to the Green function $G_0$ that solves the
KS equation:
\begin{eqnarray}
    \left[ \omega +\frac{1}{2}\nabla^{2}_{\mathbf{r}} -
    V_{\textrm{KS}}(\mathbf{r}) \right]G_0(\mathbf{r}, \mathbf{r}',
    \omega)= \delta (\mathbf{r}-\mathbf{r}' ) \label{eq:KS1},\\ V_{\textrm
    {KS}}(\mathbf{r})= V_{\textrm{H}}(\mathbf{r})+V_{\textrm {ext}}
    (\mathbf{r})+V_{\textrm{XC}}(\mathbf{r}). \label{eq:KS2}
\end{eqnarray}
The Hartree potential $V_{\textrm{H}}(\mathbf {r})=\int
\mathrm{d}\mathbf{r}^\prime v(\mathbf{r}, \mathbf{r}^\prime)
n(\mathbf{r}^\prime )$ and the exchange-correlation potential
$V_{\textrm{XC}}(\mathbf{r})$ contain the unknown density of the
interacting system, which is to be found self-consistently by the
relation $n(\mathbf{r})= -2i\int d\omega G_0(\mathbf{r},
\mathbf{r},\omega) e^{i\omega\eta} $.
The factor $e^{i\omega\eta}$, where $\eta\rightarrow0^+$, results from
time-ordering of operators and ensures appropriate analytic properties.

It is simple to check that the exchange-correlation potential in
Eq.~(\ref{eq:KS2}) modifies the Dyson equation for the Green function
into:
\begin{eqnarray}\label{eq:GWG}
&&G (\mathbf{r}_1,\mathbf{r}_2,\omega) = G_0
    (\mathbf{r}_1,\mathbf{r}_2,\omega)+\int\textrm{d}
    \mathbf{r}_3\int\textrm{d} \mathbf{r}_4 G_0
    (\mathbf{r}_1,\mathbf{r}_3,\omega) \nonumber\\&&\times \big(
    \Sigma_{\textrm{XC}} (\mathbf{r}_3,\mathbf{r}_4,\omega) -
    V_{\textrm{XC}} (\mathbf{r}_3)\delta(\mathbf{r}_3-\mathbf{r}_4)
    \big) G (\mathbf{r}_4,\mathbf{r}_2,\omega).
\end{eqnarray}
Since the Hartree potential is accounted for exactly in the KS
Eq.~(\ref{eq:KS1}), the self-energy diagrams are all of
exchange-correlation type (with no tadpoles).
The Dyson equation for the effective potential is:
\begin{eqnarray}\label{eq:GWW}
&&W (\mathbf{r}_1,\mathbf{r}_2,\omega) = v
    (\mathbf{r}_1,\mathbf{r}_2)+\nonumber\\&&\int\textrm{d}
    \mathbf{r}_3\int\textrm{d} \mathbf{r}_4 v
    (\mathbf{r}_1,\mathbf{r}_3) P (\mathbf{r}_3,\mathbf{r}_4,\omega) W
    (\mathbf{r}_4,\mathbf{r}_2,\omega).
\end{eqnarray}

Next, there are the two Hedin's equations that basically result from
the unique skeleton diagrams\cite{fet71} of the exchange-correlation
self-energy and of the polarization.
In the GWA they simplify, as one identifies the vertex with the bare
one.
\begin{eqnarray}\label{eq:GWS}
&&\Sigma_{\textrm{XC}} (\mathbf{r}_1,\mathbf{r}_2,\omega) =
    \nonumber\\&& i\int_{-\infty}^{+\infty}\frac{\textrm{d}
    \omega'}{2\pi} e^{i\omega'\eta} G
    (\mathbf{r}_1,\mathbf{r}_2,\omega+\omega') W
    (\mathbf{r}_2,\mathbf{r}_1,\omega'),
\end{eqnarray}
\begin{eqnarray}\label{eq:GWP}
    &&P (\mathbf{r}_1,\mathbf{r}_2,\omega) = \nonumber\\&&
    -2i\int_{-\infty}^{+\infty}\frac{\textrm{d}
    \omega'}{2\pi}e^{i\omega'\eta}
    G(\mathbf{r}_1,\mathbf{r}_2,\omega+\omega')
    G(\mathbf{r}_2,\mathbf{r}_1,\omega').
\end{eqnarray}

The fifth Hedin's equation is neglected in GWA and corresponds to the
skeleton expansion of the vertex, which is made of an infinite number
of diagrams.
The inclusion of vertex corrections is computationally expensive and
difficult, with a degree of arbitrariness.\cite{sch}

Finally we recall the following useful symmetry property of any
correlator due to the invariance under time-reversal of the
Hamiltonian:
\begin{eqnarray}\label{eq:prop_timerev}
    F(\mathbf{r}_1, \mathbf{r}_2,\omega )= F(\mathbf{r}_2,
    \mathbf{r}_1,\omega )
\end{eqnarray}
For the polarization $P$ and the effective potential $W$,
Eq.~(\ref{eq:prop_timerev}) implies further that they are even
functions of the frequency $\omega$.
Such exact properties are preserved by the GWA.

\section{\label{sec:method}The method}

In this section we present a method to evaluate the interacting Green
function of Eq.~(\ref{eq:GWG}) in the KS+$GW$ scheme, for an infinite
non-periodic system.
In general in the GWA one has to perform the following operations:
(i)~Evaluation of the KS Green function $G_0$ in Eq.~(\ref{eq:KS1}).
(ii)~Evaluation of the ring diagram for $P_0$ in Eq.~(\ref{eq:GWP}).
(iii)~Evaluation of $W_0$ by solving the Dyson Eq.~(\ref{eq:GWW}).
(iv)~Evaluation of $\Sigma_0=iG_0W_0$ in Eq.~(\ref{eq:GWS}).
(v)~Evaluation of $G_1$ by solving the Dyson Eq.~(\ref{eq:GWG}).

In principle, one should iterate the cycle (ii-v) for
self-consistency, by inserting $G_1$ in step~(ii).
Experience with the homogeneous electron gas (HEG) has shown that
spectral properties are better reproduced by a first-iteration
calculation rather than by a self-consistent one.\cite{hol98}

For periodic systems, the calculations can be restricted to a finite
volume (e.g., the unit cell of a lattice crystal) with suitable
boundary conditions.
The present method implements the same procedure for infinite systems in
a viable way.
A weaker hypothesis is used: along the directions of broken symmetry
any correlator $F$ can be approximated by the known $F^{\infty}$ except
for a finite length.
Hence the calculation is carried on only in a finite volume, with boundary
conditions determined by the ``asymptotic properties of the system'', i.e.,
by $F^{\infty}$.
A direct space representation will be privileged in the directions
where periodicity is absent.

\subsection{Preliminary steps}

We present two main ingredients which allow for the development of
our method: the Green function embedding method to calculate $G_0$ in
Eq.~(\ref{eq:KS1}), and a Lemma for the inversion of infinite
matrices, which permits to solve the Dyson equations for $G_1$ and
$W_0$ in Eq.~(\ref{eq:GWG}) and Eq.~(\ref{eq:GWW}), respectively.

\subsubsection{\label{sec:embedding} The embedding method }

To start, the Green function $G_0$ of the KS equation (\ref{eq:KS1})
is needed.
We make use of the Green function embedding
method.\cite{ing81,ing88,tri96}
Such a tool has been applied successfully to the study of infinite
systems without 3D periodicity, such as bulk impurities, surfaces and
adsorbates.\cite{gpb99}
Its great advantage compared to the slab and the supercell techniques
is to provide a truly continuous density of states and the correct
asymptotic behavior of all physical quantities close enough to the
defect region.
In the embedding method, space divides into a finite region $V$ and
one (or many) region $V'$ where the asymptotic regime is valid to the
required accuracy.
In this approach, the KS equation (\ref{eq:KS1}) in $V\cup V'$ is
rewritten as an equation for the finite region $V$ only.
The effect of $V'$ appears as a surface term that adds to the KS
potential.
The modified KS equation, for $\mathbf{r}$ and $\mathbf{r}'$ in $V$
reads as:
\begin{eqnarray}\label{eq:embed}
    \big[ \omega - H_{\textrm{KS}}(\mathbf{r})\big] G_0
    (\mathbf{r},\mathbf{r}',\omega ) - \nonumber \\ \int_S \mathrm{d}^2
    {\mathbf{r}}''U_S(\mathbf{r},\mathbf{r}'',\omega) G_0
    (\mathbf{r}'',\mathbf{r}',\omega )=\nonumber \\ \delta
    (\mathbf{r}-\mathbf{r}' ),
\end{eqnarray}
where $H_{\textrm{KS}}=-\frac{1}{2}\nabla_{\mathbf{r}}^{2}+V_{\textrm{KS}}$
and $S$ is the boundary of $V$.
The kernel $U_S(\mathbf{r},\mathbf{r}',\omega)$ is non-zero only for
$\mathbf{r},\mathbf{r}'{\in} S $, where it is:
\begin{eqnarray}
U_S(\mathbf{r},\mathbf{r}',\omega )=
\frac{1}{2}\delta_S(\mathbf{r},\mathbf{r}' )
\mathbf{n}_S(\mathbf{r})\cdot \nabla_\mathbf{r}+ (G^\infty_0)^{-1}
(\mathbf{r},\mathbf{r}' ,\omega).
\end{eqnarray}
$\mathbf{n}_S(\mathbf{r})$ is the unit vector normal to $S$ in
$\mathbf{r}$, pointing out of $V$.
$G^\infty_0(\mathbf{r},\mathbf{r}',\omega)$ is the Green function of
the KS equation in the asymptotic region $V'$, with Neumann
boundary condition on $S$:
\begin{eqnarray}
    [\omega - H_{\textrm{KS}} (\mathbf{r})]G_0^\infty
    (\mathbf{r},\mathbf{r}' ,\omega ) &=& \delta (\mathbf{r}- \mathbf{r}'),
    \\ {\mathbf n }_S(\mathbf{r})\cdot\nabla_{\mathbf{r}}
    G^\infty_0(\mathbf{r},\mathbf{r}' ,\omega) &=&0.
\end{eqnarray}
$(G_0^\infty)^{-1}$ is the surface functional inverse of $G_0^\infty
$, defined for points on $S$:
\begin{eqnarray}
    \int_S {\mathrm{d}}^2{\mathbf{r}}' G_0^\infty
    (\mathbf{r}_1,\mathbf{r}' ,\omega ) (G_0^\infty)^{-1}(\mathbf{r}',
    \mathbf{r}_2 ,\omega )=\delta_S(\mathbf{r}_1,\mathbf{r}_2).
\end{eqnarray}
Eq.~(\ref{eq:embed}) can be expanded over a countable basis of $V$,
thus reducing the evaluation of $G_0$ to a matrix inversion, which can
be equally done for real or complex frequencies.
We emphasize that the embedding method is formally exact so that
$G_0$ exhibits the truly {\em continuous} spectrum of the system.
Being the solution of Eq.~(\ref{eq:embed}), $G_0$ is known only for
$\mathbf{r}$ and $\mathbf{r}'$ both in $V$.
When the value of $G_0$ for one or both arguments outside $V$ is
required, it can be obtained with the ``matching Green function''
method.\cite{men03}

\subsubsection{\label{sec:lemma} Inversion of infinite matrices}

Consider the equations defining the inverse of a matrix $A$ on two
different volumes $\Omega$ and $V$, with $V\subset \Omega$:
\begin{eqnarray}
    \int_\Omega\!\mathrm{d}\mathbf{r}_3 A(\mathbf{r}_1,\mathbf{r}_3)
    A^{-1}_\Omega(\mathbf{r}_3,\mathbf{r}_2) =
    \delta(\mathbf{r}_1-\mathbf{r}_2), \; \mathbf{r}_1,\mathbf{r}_2
    \in \Omega, \label{Lem1}\\ \int_V\!\mathrm{d}\mathbf{r}_3
    A(\mathbf{r}_1,\mathbf{r}_3) A^{-1}_V(\mathbf{r}_3,\mathbf{r}_2) =
    \delta(\mathbf{r}_1-\mathbf{r}_2), \; \mathbf{r}_1,\mathbf{r}_2
    \in V.
    \label{Lem2}
\end{eqnarray}
In general $A^{-1}_V$ is different from the restriction of
$A^{-1}_\Omega$ in $V$.
However, the following Lemma gives a condition for the two matrices to
coincide on a smaller subset $U\subset V\subset \Omega $.\\
{\textbf {Lemma}}: If $ A_\Omega^{-1}(\mathbf{r}_1,\mathbf{r}_2)=0 $
for all $\mathbf{r}_1\in \Omega-V$ and $\mathbf{r}_2\in U$, then
$A_\Omega^{-1}(\mathbf{r}_1,\mathbf{r}_2)=A_V^{-1}(\mathbf{r}_1,\mathbf{r}_2)$
for all $\mathbf{r}_1,\mathbf{r}_2\in U$.\\
Proof: consider Eq.~(\ref{Lem1}) for $\mathbf{r}_1\in V$ and
$\mathbf{r}_2\in U$, multiply it by
$A_V^{-1}(\mathbf{r}_4,\mathbf{r}_1)$ and integrate in $\mathbf{r}_1$
over $V$:
$$\int_V \!\!\mathrm{d}\mathbf{r}_1
A_V^{-1}(\mathbf{r}_4,\mathbf{r}_1)
\!\!\int_\Omega \!\! \mathrm{d}\mathbf{r}_3
A (\mathbf{r}_1,\mathbf{r}_3) A_\Omega^{-1}(\mathbf{r}_3,\mathbf{r}_2)
= A_V^{-1}(\mathbf{r}_4,\mathbf{r}_2).$$
The integral in $\mathbf{r}_3$ over $\Omega$ is split into an integral
on $V$ and on $\Omega -V$.
The first integral yields $A_\Omega^{-1}(\mathbf{r}_4,\mathbf{r}_2)$,
the latter vanishes because
$A_\Omega^{-1}(\mathbf{r}_1,\mathbf{r}_2)=0$ for $\mathbf{r}_1\in
\Omega -V$ and $\mathbf{r}_2\in U$. $\bullet $\\
As a consequence of the Lemma, if we are interested in the values of 
$A_\Omega^{-1}$ in a subset $U$ of the possibly infinite volume $\Omega$,
it is sufficient to invert $A$ on a suitable {\em larger} subset $V$, with
$V\subset \Omega$.
Quite generally, the functions of interest have the property
$A_\Omega^{-1}(\mathbf{r}_1,\mathbf{r}_2)\rightarrow0$ as
$|\mathbf{r}_1-\mathbf{r}_2|\rightarrow\infty$.
Therefore, the hypothesis of the Lemma can be regarded as true to any
degree of accuracy, for a large enough set $V$.

\subsection{\label{sec:met_steps}Steps of the method}

\subsubsection{\label{sec:met_p}The polarization}

In direct space and GWA, the polarization $P_0$ is given by the ring
diagram [Eq.~(\ref{eq:GWP})] with Green functions $G_0$.

To compute it numerically, we note that the convergence of the
integration in Eq.~(\ref{eq:GWP}) is improved by using $\Delta
G_0=G_0-G_0^{\infty}$, which decays faster than $G_0$ as
$|\omega|\rightarrow\infty$.
Hence we can take advantage
of the known function $P_0^{\infty}$, that corresponds to the
asymptotic limit of $P_0$ continued into the region of interest.
We can write:
\begin{eqnarray}\label{eq:GWP0}
    P_0(\omega ) &=& \int \mathrm{d}\omega' \big[G_0^{\infty}+\Delta
    G_0\big] (\omega+\omega') \nonumber\\ &&\times\big[G_0^{\infty}+\Delta
    G_0\big](\omega').
\end{eqnarray}
The integration of the term containing ($G_0^\infty G_0^\infty$) gives
$P_0^{\infty} $, the others have a faster decay for large $|\omega|$.
Owing to the non-analytic behaviour of the Green function close to the real
axis, it is convenient to compute Eq.~\ref{eq:GWP0} on the contour described
in the Appendix.
Since the spatial dependence is not involved in the computation of the
polarization $P_0({\mathbf{r}}_1,{\mathbf{r}}_2,\omega)$, such
dependence amounts to that of $G_0$ in the same region.

\subsubsection{\label{sec:met_w}The effective potential}

The effective (dressed) potential is the solution of the Dyson
equation, Eq.~(\ref{eq:GWW}), and it is formally given by the integral:
\begin{eqnarray}\label{eq:mw}
    W_0(\mathbf{r}_1,\mathbf{r}_2,\omega )= \int \mathrm{d}
    \mathbf{r}_3 \varepsilon^{-1} (\mathbf{r}_1,\mathbf{r}_3,\omega)
    v(\mathbf{r}_3,\mathbf{r}_2).
\end{eqnarray}
where $\varepsilon^{-1}$ is the functional inverse of the dielectric
function $\varepsilon$:
\begin{eqnarray}\label{eq:mepsilon}
    \varepsilon (\mathbf{r}_1,\mathbf{r}_2,\omega)&=& \delta
    (\mathbf{r}_1-\mathbf{r}_2)\nonumber\\
    &&-\int\mathrm{d}\mathbf{r}_3 v(\mathbf{r}_1,\mathbf{r}_3)
    P_0(\mathbf{r}_3,\mathbf{r}_2,\omega).
\end{eqnarray}
The decay properties of the polarization $P_0$ as
$|\mathbf{r}_1-\mathbf{r}_2|\rightarrow\infty$ imply that
Eq.~(\ref{eq:mepsilon}) can be evaluated numerically by introducing
cutoffs for the integration variable $\mathbf{r}_3$.

The inverse dielectric function is formally defined by:
\begin{eqnarray}\label{eq:inveps}
    \int \mathrm{d}\mathbf{r}_3 \,\varepsilon
    (\mathbf{r}_1,\mathbf{r}_3,\omega) \varepsilon^{-1}
    (\mathbf{r}_3,\mathbf{r}_2,\omega) = \delta
    (\mathbf{r}_1-\mathbf{r}_2).
\end{eqnarray}
The inversion of the matrix $\varepsilon$ on an unbounded region is
not a feasible numerical calculation.
However, we only need evaluate $\varepsilon^{-1}$ on a finite region
$U_\varepsilon $ outside which the asymptotic regime holds.
Since in ordinary systems the effective interaction between electrons
far apart decays to zero as the distance increases, Eq.~(\ref{eq:mw})
implies that also $\varepsilon^{-1}(\mathbf{r}_1,\mathbf{r}_2)$ goes
to zero as $|\mathbf{r}_1-\mathbf{r}_2|\rightarrow\infty$.
So we can use the Lemma proven in Sec.~\ref{sec:lemma} and restrict
the integration in Eq.~(\ref{eq:inveps}) to a finite region
$V_\varepsilon $, $U_\varepsilon \subset V_\varepsilon$, still
obtaining correct values of
$\varepsilon^{-1}(\mathbf{r}_1,\mathbf{r}_2)$ for
$\mathbf{r}_1,\mathbf{r}_2\in{}U_\varepsilon $.
An expansion over a discrete basis set is now possible, leading the
problem to an ordinary matrix inversion.
The size of the region $V_\varepsilon $ is simply a numerical
parameter, and convergence in the resulting $\varepsilon^{-1}$ must be
checked.
With this purpose, a localized basis set [like
$\phi_i(\mathbf{r})=c\delta(\mathbf{r}-\mathbf{r}_i)$] is more
convenient.
In this case the matrix elements $\varepsilon^{-1}_{ij}$, being
directly proportional to
$\varepsilon^{-1}(\mathbf{r}_i,\mathbf{r}_j)$, do not change except
for the normalization factor $c$, when the size of $V_\varepsilon$ or
the number of basis function changes.
The effective potential can now be evaluated from Eq.~(\ref{eq:mw}).
We only remark that the frequency argument $\omega$ is fixed, so that
$W_0$ is computed for the same values of $\omega$ as those chosen for the
polarization -- $\omega$ purely imaginary as shown in the Appendix.

Finally we note that the assertion that
$\varepsilon^{-1}(\mathbf{r}_1,\mathbf{r}_2)\rightarrow0$ as
$|\mathbf{r}_1-\mathbf{r}_2|\rightarrow\infty$ {\em before}
$\varepsilon^{-1}$ is actually evaluated does not pose conceptual
difficulties:
the decay to zero has in fact to be checked for the known function
$(\varepsilon^{\infty})^{-1}$.

\subsubsection{\label{sec:met_s} The self-energy}

It is customary to split the self-energy into the sum of the exchange
$\Sigma_{\textrm{X}}=iGv$ and the correlation term
$\Sigma_{\textrm{C}}=iG(W-v)$.
The evaluation of the exchange term poses no problem:
\begin{eqnarray}\label{eq:GWSX}
\Sigma_{\textrm{X}} (\mathbf{r}_1,\mathbf{r}_2) = i
    v(\mathbf{r}_1,\mathbf{r}_2)
    \int_{-\infty}^{+\infty}\frac{\textrm{d} \omega}{2\pi}
    e^{i\omega\eta} G_0(\mathbf{r}_1,\mathbf{r}_2,\omega).
\end{eqnarray}
The frequency integration sums over the occupied KS states, and can be
performed analytically if the spectrum is discrete.
In our case, the spectrum is generally continuous [$G_0$ is the
solution of Eq.(\ref{eq:embed})], and the sum is replaced by an
integral to be performed numerically.
A change of contour proves useful, together with the information that
we shall be dealing with systems in which there are no KS states below
the bottom of the band $E_0$ ($\eta'$ is a positive infinitesimal):
\begin{eqnarray}
    &&\int_{-\infty}^{+\infty}\textrm{d} \omega e^{i\omega\eta}
    G_0(\omega)\nonumber\\ &&=\int_{E_0}^{\mu}\textrm{d} \omega
    \big[G_0(\omega-i\eta') - G_0(\omega+i\eta') \big].
\end{eqnarray}
The resulting integral is computed straightforwardly.
$\mu$ is the KS chemical potential.

The correlation term is:
\begin{eqnarray}\label{eq:GWSC}
    \Sigma_{\textrm{C}} (\omega)=i\int_{-\infty}^{+\infty}
    \mathrm{d}\omega' e^{i\omega'\eta} G_0(\omega
    +\omega')\big[W_0(\omega' )-v\big].
\end{eqnarray}
Since the main contribution to the above integral is
$\Sigma_{\textrm{C}}^{\infty}$, one can use the same splitting of the
self-energy in Eq.~(\ref{eq:GWSC}) into a leading asymptotic term plus
a correction due to the inhomogeneity as in Sec.~\ref{sec:met_p} for
the polarization.

Mathematical details on the calculation of the integrals determining
the the self-energy are discussed in the Appendix.

\subsubsection{\label{sec:met_g} The Green function}

The Dyson equation for $G$ [Eq.~(\ref{eq:GWG})] is formally the same
as the Dyson equation for $W$ [Eq.~(\ref{eq:GWW})] once $G$,
$\Sigma_{\textrm{XC}}-V_{\textrm{XC}}$ and $G_0$ are identified with
$W$, $P$, and $v$ respectively.
Therefore we define the function
$\epsilon_{\textrm{XC}}(\mathbf{r}_1,\mathbf{r}_2,\omega)$, which is
analogous to the dielectric function:
\begin{eqnarray}\label{eq:m_g}
    &&\epsilon_{\textrm{XC}}(\mathbf{r}_1,\mathbf{r}_2,\omega) =
     \delta(\mathbf{r}_1-\mathbf{r}_2) -\int\mathrm{d}\mathbf{r}_3
     G_0(\mathbf{r}_1,\mathbf{r}_3,\omega) \nonumber\\&&\times
     \big[\Sigma_{\textrm{XC}}(\mathbf{r}_3,\mathbf{r}_2,\omega)-
     V_{\textrm{XC}}(\mathbf{r}_2)\delta
     (\mathbf{r}_3-\mathbf{r}_2)\big].
\end{eqnarray}

Even with the these strong analogies, $\epsilon_{\textrm{XC}}$
presents a striking difference with $\varepsilon$:
the decay of $\epsilon_{\textrm{XC}}(\mathbf{r}_1,\mathbf{r}_2)$ as
$|\mathbf{r}_1,\mathbf{r}_2|$ goes to infinity
is linked to the one of $G_0$,
which in turn varies according to the dimensionality of the system.
As a consequence, $\epsilon_{\textrm{XC}}(\mathbf{r}_1,\mathbf{r}_2)$ may not
go to zero as $|\mathbf{r}_1-\mathbf{r}_2|$ goes to infinity, thus not
satisfying the hypothesis of the Lemma in Sec.~\ref{sec:lemma}.
To make $G_0$ decay as $|\mathbf{r}-\mathbf{r}'|\rightarrow\infty$
(which implies the same property for $\epsilon_{\textrm{XC}}$), it is
convenient to solve Eq.~(\ref{eq:m_g}) at a complex frequency
$\omega+i\Delta$.
The choice of the real quantity $\Delta$ depends on a compromise: if
it is too small, the decay of $G_0$ is very slow and a large region of
inversion of $\epsilon_{\textrm{XC}}$ is needed; if it is too large,
the structures on the real frequency axis we are interested in are
broadened -- in fact $\Delta$ plays the role of resolution.
The final result of the calculation, the interacting Green function,
is thus evaluated on a translated frequency axis $\omega+i\Delta$.
Analytical continuation improves the resolution: first, the values of
$G$ are fitted with a rational function, then the expression is
continued to the real frequency axis.

\section{\label{sec:semiinfj}Semi-infinite jellium}

In this section we illustrate the application of the method to
semi-infinite jellium.\cite{dob95}

\subsection{Introduction to the system}

Semi-infinite jellium is a neutral system of dynamical electrons in a
background of uniform positive charge density in the half-space $z\leq0$.
In the half-space $z>0$ there is no positive charge.
We denote the uniform density by $n$, $n=1/(4\pi{}r_{s}^{3}/3)$.
In the computations the value of $r_s$ will be fixed to give an
electron density equal to that of aluminium ($r_s=2.07$~$a_0$).

Besides time translations and time reversal, the system is also
invariant under translations parallel to the surface.
Hence the wave-vector parallel to the surface ${\mathbf{k}}_\parallel$ is
a good quantum number.
It is important to observe that since the solid is semi-infinite,
the wave-vector $k_z$ may take any real value.
So we shall deal with truly continuous densities of states, also for a
fixed ${\mathbf{k}}_\parallel$.
A slab calculation would instead determine a discrete spectrum (i.e.,
a nonphysical quantized set of $k_z$ wave-vectors).
Because of the just mentioned symmetry, a two-point correlator
$F(\mathbf{r}_\parallel,z,;\mathbf{r}'_\parallel,z';\omega)$ only
depends on the difference
$\mathbf{r}_\parallel-\mathbf{r}'_\parallel$, where
$\mathbf{r}_\parallel=(x,y)$.
We can express $F$ in terms of its Fourier transform (FT) with respect
to a wave vector parallel to the surface:
\begin{eqnarray}
    F(\mathbf{r}_\parallel,z;\mathbf{r}'_\parallel,z';\omega) \! =
    \!\!\int\!\!
    \frac{\textrm{d}\mathbf{k}_\parallel}{(2\pi)^2} 
    e^{i\mathbf{k}_\parallel\cdot(\mathbf{r}_\parallel-\mathbf{r}'_\parallel)}
    F(z,z',\mathbf{k}_\parallel,\omega).
\end{eqnarray}
Since the system is also invariant under rotations around the
$z$-axis, $F$ depends only on the modulus of $\mathbf{k}_\parallel$.

The study of semi-infinite jellium is basically one-dimensional and
satisfies the requirements of applicability of our method.
In fact, the perturbation induced by the jellium edge is localized
near the surface.
At a distance of few $r_s$ inside the solid, the properties of the
system approach those of the infinite, homogeneous electron gas (HEG).
Many-body results for the HEG in the $GW$ approximation are well known
in the literature.\cite{hol98}
If we indicate by $F^{\textrm{HEG}}_{n}(k,\omega)$ the correlator $F$
evaluated for a HEG of density $n$, the bulk and vacuum limits of $F$
are
\begin{eqnarray}\label{eq:jFB}
    F_{\textrm{B,V}}^{\infty}(z,z',k_\parallel,\omega) \!\!=
    \!\!\int\!\frac{\mathrm{d}k_z}{2\pi} \! e^{ik_z(z-z')}
    F^{\textrm{HEG}}_{n,0}(\sqrt{
    \mathbf{k}_\parallel^2+k_{z}^{2}},\omega).
\end{eqnarray}

Therefore we need evaluate the correlator $F$ only on a limited
interval $[z_{\textrm{B}},z_{\textrm{V}}]$ of the $z$-axis, which will
be the volume $U_F$ of the previous Sec.~\ref{sec:lemma}.

Care must be taken when $F$ is evaluated at $z$ and $z'$ in different
regions of space, e.g., $z$ in bulk and $z'$ in vacuum.
However, we are working with functions with the important property
$F(z,z')\rightarrow0$ when $|z-z'|\rightarrow\infty$.
This guarantees that when $z$ is in bulk and $z'$ in vacuum the
functions $F_{\textrm{B}}(z,z')$ and $F_{\textrm{V}}(z,z')$ are both
zero to the desired accuracy, provided that
$|z_{\textrm{B}}-z_{\textrm{V}}|$ is large enough.

The dependence of $F$ on its four arguments
$(z,z',k_\parallel,\omega)$ is in principle continuous.
Numerically, we stored the information about $F$ on a four dimensional
discrete mesh.
Intermediate values, when necessary, are obtained with interpolation
algorithms.
For $z$ and $z'$ meshes, natural limit values are given by
$z_{\textrm{B}}$ and $z_{\textrm{V}}$.
Cutoffs for $k_\parallel$ and $\omega$ can be fixed since
$F\rightarrow0$ when $k_\parallel\rightarrow\infty$ or
$|\omega|\rightarrow\infty$.
Different meshes have to be chosen for different functions.
Numerical convergence must be checked for any parameter defining the
mesh.

\subsection{$G_0W_0$ equations}

In the chosen representation, the $G_0W_0$ equations
(\ref{eq:GWG}-\ref{eq:GWP}) take the form:
\begin{widetext}
\begin{subequations}
\label{eq:jGW}
\begin{eqnarray}\label{eq:jGWP}
    P_0 (z_1,z_2,k_\parallel,\omega) =
    -2i\int_{-\infty}^{+\infty}\frac{\textrm{d} \omega'}{2\pi}
    \int\frac{\textrm{d}^2\mathbf{k}'_\parallel}{(2\pi)^2}
    e^{i\omega'\eta}
    G_0(z_1,z_2,|\mathbf{k}_\parallel+\mathbf{k}'_\parallel|,\omega+\omega')
    G_0(z_2,z_1,k'_\parallel,\omega'),
\end{eqnarray}
\begin{eqnarray}\label{eq:jGWW}
W_0 (z_1,z_2,k_\parallel,\omega) = v
    (|z_1-z_2|,k_\parallel)+\int\textrm{d} z_3\int\textrm{d} z_4 v
    (|z_1-z_3|,k_\parallel) P_0 (z_3,z_4,k_\parallel,\omega) W_0
    (z_4,z_2,k_\parallel,\omega),
\end{eqnarray}
\begin{eqnarray}\label{eq:jGWS}
\Sigma_{\textrm{XC}} (z_1,z_2,k_\parallel,\omega) =
    i\int_{-\infty}^{+\infty}\frac{\textrm{d} \omega'}{2\pi}
    \int\frac{\textrm{d}^2\mathbf{k}'_\parallel}{(2\pi)^2}
    e^{i\omega'\eta}
    G_0(z_1,z_2,|\mathbf{k}_\parallel+\mathbf{k}'_\parallel|,\omega+\omega')
    W_0(z_2,z_1,k'_\parallel,\omega'),
\end{eqnarray}
\begin{eqnarray}\label{eq:jGWG}
G_1 (z_1,z_2,k_\parallel,\omega) &=& G_0
    (z_1,z_2,k_\parallel,\omega)+\int\textrm{d} z_3\int\textrm{d} z_4
    G_0 (z_1,z_3,k_\parallel,\omega) \nonumber\\&&\times \big(
    \Sigma_{\textrm{XC}} (z_3,z_4,k_\parallel,\omega) -
    V_{\textrm{XC}} (z_3)\delta(z_3-z_4) \big) G_1
    (z_4,z_2,k_\parallel,\omega),
\end{eqnarray}
\end{subequations}
\end{widetext}
where $v$ is the FT of Coulomb's potential:
\begin{eqnarray}
v (|z_1-z_2|,k_\parallel) =
    \frac{2\pi}{k_\parallel}e^{-k_\parallel|z_1-z_2|}.
\end{eqnarray}

\subsection{Steps of the calculation and results}

We omit details regarding the starting point: the KS Green function
$G_0(z,z',k_\parallel,\omega)$ can be obtained straightforwardly, with
the embedding method described in Sec.~\ref{sec:embedding}.
We calculated it in the LDA.

\subsubsection{The polarization}

The parallel-wavector convolution in Eq.~(\ref{eq:jGWP}) does not pose
numerical difficulties.
Regarding the frequency convolution, the factor $e^{i\omega'\eta}$ is
necessary for the convergence of the integral.
In fact, the Green function $G_0(\omega)$ approaches zero like
$|\omega|^{-1/2}$ as $|\omega|\rightarrow\infty$ (this can be easily
verified for the HEG Green function).
We follow the treatment of Sec.~\ref{sec:met_p}, and separate the integrand
in two terms: one
($G^{\textrm{HEG}}_{n} G^{\textrm{HEG}}_{n}$), that gives
$P^{\textrm{HEG}}_{n}$, and ($G_0 G_0 - G^{\textrm{HEG}}_{n}
G^{\textrm{HEG}}_{n}$) that can be evaluated numerically, because it
goes to zero as $|\omega|^{-3/2}$ when $|\omega|\rightarrow\infty$.

\subsubsection{The effective potential}

As shown in Sec.~\ref{sec:met_w}, one has to evaluate the dielectric
function $\varepsilon$, defined by
\begin{eqnarray}\label{eq:jeps}
    && \varepsilon(z_1,z_2,k_\parallel,\omega) = \delta(z_1-z_2)\\ &&
    -\int\mathrm{d}z_3v(|z_1-z_3|,k_\parallel)
    P_0(z_3,z_2,k_\parallel,\omega),\nonumber
\end{eqnarray}
and its inverse $\varepsilon^{-1}$ in order to solve Dyson equation
for $W$, Eq.~(\ref{eq:jGWW}):
\begin{eqnarray}\label{eq:jeps-1v}
    &&W_0(z_1,z_2,k_\parallel,\omega)=\nonumber\\&&\int
    \mathrm{d}z_3
    \,\varepsilon^{-1}(z_1,z_3,k_\parallel,\omega)v(|z_3-z_2|,k_\parallel).
\end{eqnarray}
The $z$ integrations may be restricted to finite intervals since
$P_0(z_1,z_2)$ and $\epsilon^{-1}(z_1,z_2)$ go to $0$ as
$|z_1-z_2|\rightarrow\infty$.
We also point out that the calculation time required by the integrals
in Eqs.~(\ref{eq:jeps}) and~(\ref{eq:jeps-1v}) is negligible, so the
cutoffs of $z$ can be cautiously overestimated.

To obtain $\varepsilon^{-1}$, defined by the inversion over the whole
$z$-axis,
\begin{eqnarray}\label{eq:jinveps}
    \int_{-\infty}^{+\infty}\mathrm{d}{z_3}
    \varepsilon^{-1}(z_1,z_3)\varepsilon(z_3,z_2) = \delta(z_1-z_2),
\end{eqnarray}
on the finite interval
$U_\varepsilon=[z_{\textrm{B}},z_{\textrm{V}}]$, we exploit the Lemma
in Sec.~\ref{sec:lemma} and restrict the integration in
Eq.~(\ref{eq:jinveps}) to the finite interval
$V_\varepsilon=[L_{\textrm{B}},L_{\textrm{V}}]$, $U_\varepsilon\subset
V_\varepsilon$.

\begin{figure}
    \includegraphics[width=1.0\columnwidth]{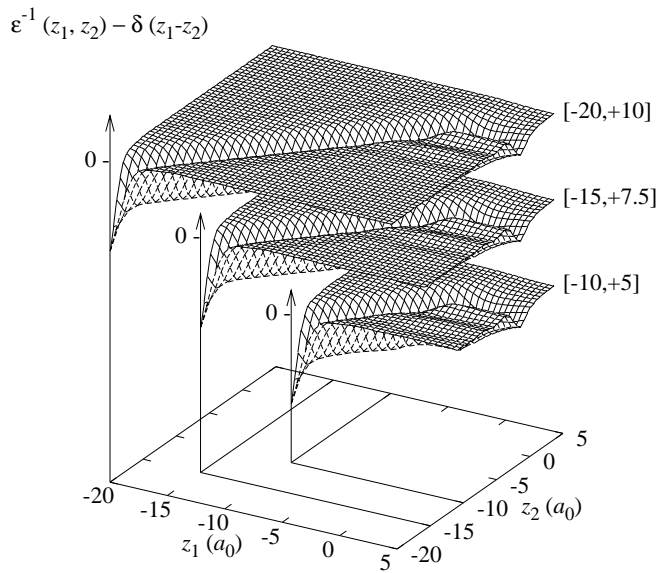}
    \caption{\label{fig:jinveps} Values of
    $\varepsilon^{-1}(z_1,z_2)-\delta(z_1-z_2)$ for
    different inverting regions
    $V_\varepsilon=[L_{\textrm{B}},L_{\textrm{V}}]$. Here,
    $r_s=2.07$~$a_0$, $\omega=0$ and $k_\parallel=0.2k_F$, $k_F$ being the
    Fermi wave-vector.}
\end{figure}

Correct values of $\varepsilon^{-1}(z_1,z_2)$ for
$z_1,z_2\in{}U_\varepsilon $ are obtained if $L_{\textrm{B}}$ and
$L_{\textrm{V}}$ are conveniently large.
It is sufficient that
$\varepsilon^{-1}(z,z_{\textrm{B}})\simeq0$ if $z<L_{\textrm{B}}$ and
$\varepsilon^{-1}(z,z_{\textrm{V}})\simeq0$ if $z>L_{\textrm{V}}$
to the required degree of accuracy.
This concept is graphically presented in Fig.~\ref{fig:jinveps}, where
we show $\varepsilon^{-1}$ for different values of $L_{\textrm{B}}$
and $L_{\textrm{V}}$.
The negative peak for $z_1,z_2=L_{\textrm{B}}$ (on the left of each
plot) represents a spurious feature introduced when the region of integration
is restricted to a finite interval.
Notice that this behavior is located at the boundaries of
$V_\varepsilon$ regardless to its size.
Since $\varepsilon^{-1}(z_1,z_2)$ is different from zero only for
values of $z_2$ close to $z_1$, this nonphysical feature is confined
within few atomic units from the boundaries.
Hence the interval $V_\varepsilon$ has to be only slightly larger than
$U_\varepsilon $.
For the values described in Fig.~\ref{fig:jinveps}, if
$z_{\textrm{B}}=-15$~$a_0$, the choice $L_{\textrm{B}}=-20$~$a_0$ is
already an accurate one.
A similar discussion has to be done with respect to $L_{\textrm{V}}$,
but in this case the spurious peak is much smaller.

\begin{figure}
    \includegraphics[width=1.0\columnwidth]{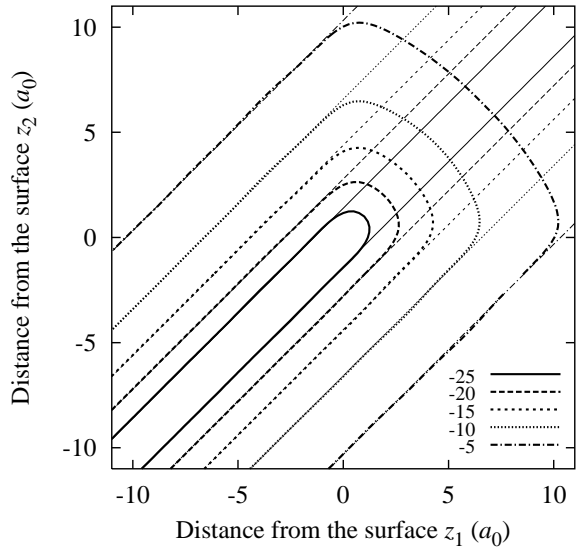}
    \caption{\label{fig:wconbulk} Contour levels of
    $W_0(z_1,z_2,k_\parallel,\omega)-v(|z_1-z_2|,k_\parallel)$ in the
    $z_1,z_2$ plane. Thick curves: semi-infinite jellium. Thin curves: HEG
    of equal density. Here $r_s=2.07$~$a_0$, $\omega=0$ and
    $k_\parallel=0.2k_F$.}
\end{figure}

In Fig.~\ref{fig:wconbulk} we show the contour levels of the difference
between the effective and bare interaction $W_0-v$ in the $z_1,z_2$ plane.
The HEG levels of the same quantity are also reported.
The agreement is excellent when $z_1$ and $z_2$ approach bulk.
As we move into the vacuum, $W_0-v$ correctly goes to zero.

\begin{figure}
    \includegraphics[width=1.0\columnwidth]{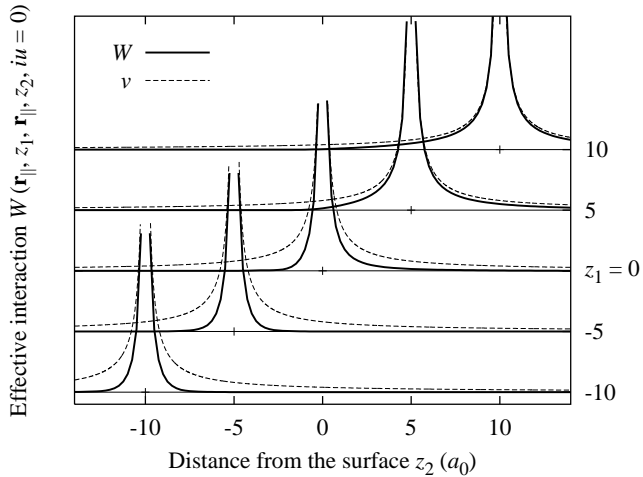}
    \caption{\label{fig:wrr} Effective and bare Coulomb interactions near
    the jellium surface for points aligned on the normal.
    $r_s=2.07$~$a_0$.}
\end{figure}

Next we consider the effective potential $W_0$ in the more intuitive
representation, $(\mathbf{r}_1,\mathbf{r}_2,\omega)$, obtained by
anti-FT with respect to $\mathbf{k}_\parallel$.
For simplicity, we limit our discussion to the static case
($\omega=0$) and consider collinear points on the normal to the
surface ($\mathbf{r}_{1\parallel}=\mathbf{r}_{2\parallel}$).
Fig.~\ref{fig:wrr} shows the effective potential as a function of $z$
from bulk to vacuum: $W_0$ is similar to a Yukawa screened potential for
$z_1$ and $z_2$ in bulk, and it coincides with the bare Coulomb
interaction for $z_1$ and $z_2$ in vacuum.
Some intermediate values are shown: for $z_1$ fixed near the surface,
$W_0$ is no longer a symmetric function of $z_2$ with respect to
$z_1$, being the screening inhomogeneous.

\subsubsection{The self-energy}

The evaluation of the self-energy follows closely that of the
polarization.
However, differently from the previous case, it is not necessary to
split the self-energy $\Sigma_{\textrm{C}}$ into a leading term and a
correction to it, as discussed in Sec.~\ref{sec:met_s}, because the
difference $W-v$ decays fast enough to ensure convergence.

\begin{figure}
    \includegraphics[width=1.0\columnwidth]{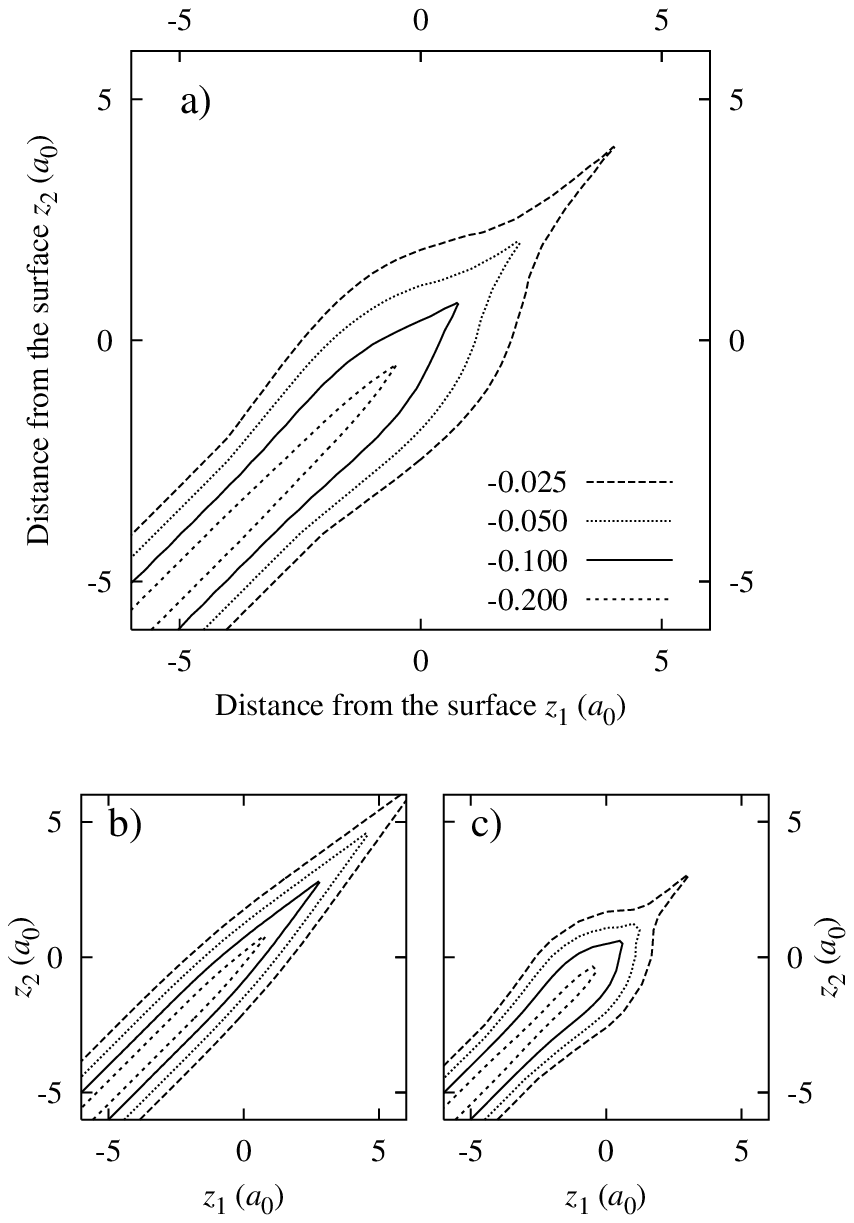}
    \caption{\label{fig:sigmazz} Self-energy
    $\Sigma_{\textrm{XC}}(z_1,z_2,k_\parallel,\omega)$ for
    $r_s=2.07$~$a_0$, $k_\parallel=0$ and $\omega=\mu$. (a)~Result of the
    GW computation. (b)~Paula Sanch\'ez-Friera's model.
    (c)~Step-potential approximation. Contour levels as in (a).}
\end{figure}

In Fig.~\ref{fig:sigmazz} we report the contour levels of the
self-energy evaluated at $k_\parallel=0$ and $\omega=\mu$ .
Fig.~\ref{fig:sigmazz}(a) shows the result of our calculation: a
particular feature in the near-surface region is the ``arabian'' shape
of the contours levels.
This arises because when $z_1$ and $z_2$ lie outside jellium, the
self-energy, as $|z_1-z_2|$ increases, decreases in a slower way than
in bulk owing to lower screening.

In order to reduce the numerical effort required by a $GW$
calculation, attempts have been made to mimic the self-energy with
efficient models.
In Fig.~\ref{fig:sigmazz}(b) we report the self-energy of a model
based on fitting the spatial dependence of $\Sigma$ from that of the
HEG having the average density of the system, successfully tested for
bulk  silicon.\cite{san00}
The surface peculiar contours in Fig.~\ref{fig:sigmazz}(a) cannot be
reproduced by this model and the decay of $\Sigma$ in vacuum is much slower.
The comparison between these two results points out that a model based
on an average density cannot reproduce features which are
characteristic of a surface.

In Fig.~\ref{fig:sigmazz}(c) we show the self-energy evaluated from a
$G_0W_0$ calculation where, instead of the self-consistent DFT-LDA
Green function $G_0$, one uses the Green function
$G^{\textrm{step}}_0$ computed for the approximated effective
potential:
$v_{\textrm{eff}}^{\textrm{step}}(z)=v_{\textrm{XC}}(-\infty)$ if $z\leq0$
and $v_{\textrm{eff}}^{\textrm{step}}(z)=0$ otherwise.
It is well know that in this case there are stronger Friedel's oscillations,
and the electronic states (and thus the Green function) decay
faster away from the surface.
This is reflected in the self-energy.
Finally we remark that the difference between $G_0W_0$ and
$G^{\textrm{step}}_0W^{\textrm{step}}_0$ is due primarily to the
difference between the Green functions, whereas the effective
interactions are very similar.

\subsubsection{The interacting Green function}

Following Eq.~(\ref{eq:m_g}), the interacting Green function $G_1$ for
the semi-infinite jellium can be evaluated from
\begin{eqnarray}\label{eq:jepsxc-1G_0}
    &&G_1(z_1,z_2,k_\parallel,\omega)=\nonumber\\&&\int
    \mathrm{d}z_3\,\epsilon^{-1}_{\textrm{XC}}(z_1,z_3,k_\parallel,\omega)
    G_0(z_3,z_2,k_\parallel,\omega),
\end{eqnarray}
where the kernel $\epsilon^{-1}_{\textrm{XC}}$ is obtained by inverting
\begin{eqnarray}\label{eq:jepsxc}
    &&\epsilon_{\textrm{XC}}(z_1,z_2,k_\parallel,\omega) =
    \delta(z_1-z_2) -\int\mathrm{d}z_3G_0(z_1,z_3,k_\parallel,\omega)
    \nonumber\\&&\times
    \big[\Sigma_{\textrm{XC}}(z_3,z_2,k_\parallel,\omega)-
    V_{\textrm{XC}}(z_2)\delta(z_3-z_2)\big].
\end{eqnarray}

When the frequency is infinitesimally close to the real axis and its
real part corresponds to extended states,
$G_0(z_1,z_3,k_\parallel,\omega)$, as a function of $z_3$, is a plane
wave with undamped oscillations propagating to infinity.
As a consequence $\epsilon^{-1}_{\textrm{XC}}(z_1,z_2)$ does not go to
zero as $|z_1-z_2|$ goes to infinity, thus not satisfying the
hypothesis of the Lemma.
One must then recur to the procedure outlined in Sec.~\ref{sec:met_g}, i.e.,
solve Eq.~(\ref{eq:jGWG}) at a complex frequency $\omega+i\Delta$.
We experienced that an interval $V_\epsilon$ about $100$~$a_0$ wide
was needed for a value of $\Delta$ of about $0.05$~hartree, in order
to describe the surface region correctly.

The imaginary part of the interacting Green function $G$ yields the
many-body spectral weight function
\begin{eqnarray}\label{eq:A}
    A(z,\mathbf{k}_\parallel,\omega)=-\frac{1}{\pi}
    \Im{}G(z,z,\mathbf{k}_\parallel,\omega) \textrm{sgn}(\omega-\mu).
\end{eqnarray}
This quantity gives a measure of the quasi-particle amplitude and is
directly related to a variety of experiments such as photoemission
spectroscopies,\cite{mat98} and scanning tunneling
microscopy.\cite{sch01}
The integral in $\mathbf{k}_\parallel$ gives the local density of
states (LDOS)
\begin{eqnarray}\label{eq:ldosA}
    \sigma(z,\omega)=\int\frac{\mathrm{d}^2
    \mathbf{k}_\parallel}{(2\pi)^2} A(z,\mathbf{k}_\parallel,\omega).
\end{eqnarray}

The evaluation of the LDOS of semi-infinite jellium in this framework
demonstrates the presence at the surface of a broad image-potential
induced (IPI) resonance, which emerges sharply when results are
compared to DFT-LDA ones.
We stress that an IPI resonance width can only be obtained by a many-body
approach like ours which takes into account the semi-infinite character of
the solid.
We refer to Ref.~\onlinecite{fra03} for the results and a detailed
discussion on this topic.

\section{\label{sec:concl}Conclusions}

We have presented a method to investigate infinite non-periodic
systems in the framework of the GWA. 
Calculations can be performed in finite regions, without introducing
nonphysical boundary conditions,
such as confining barriers (the slab approach) or a 3D fictitious
periodicity (the supercell one).
In such systems (e.g., a solid with a surface) densities of states are
continuous, and while really discrete states may exist inside gaps,
other ones become resonances when they do overlap in energy with a
continuum band.
The proposed method is particularly suitable for the description of
these systems.
In fact on the one side the embedding approach, which allows for
calculating a truly continuous density of states, includes
automatically the hybridization between bulk and surface states.
On the other many-body effects, whose treatment is needed for excited
states or image potential ones, are accounted for at the GWA level.

On the contrary a DFT slab calculation of such systems (e.g., in the
LDA or GGA) is only able to work out a spectral weight constituted by
delta functions, one for each discrete eigenstate, while the real
structure of the spectrum may be in general more complicate as just
outlined.
The GWA correction cannot amend by itself this result, but only
determine a broadening of quasi-particle states (plus eventually minor
additional structures) due to many-body correlations.
This broadening, which can be evaluated in first approximation by
taking the average value of the self-energy over the DFT state, may be
much smaller than that due to hybridization  effects, as it is the
case for IPI resonances.

In this paper we have also extensively investigated semi-infinite
jellium by our approach.
We have illustrated the bulk-to-vacuum transition of the many-body
electron gas properties.
By comparing the LDA and GWA density of states, this method has been able
to identify an image potential surface resonance of large
width.\cite{fra03}
Extension of this approach to semi-infinite realistic surfaces could
bring a wealth of accurate data on the spectral properties of surfaces
and adsorbates, especially regarding the excited states.

\begin{acknowledgments}
We would like to thank M.~I.~Trioni and G.~Onida for useful
discussions.
This work was supported by the Italian MIUR through Grant
No. 2001021128.
\end{acknowledgments}

\appendix*\section{\label{appendix}Analytic continuation of frequency
integrals.}

The presence of non-analyticities close to the contour of the frequency
integration renders it difficult to integrate expressions containing
the Green function $G$ and the effective potential $W$ numerically, as
for the polarization or the self-energy.

Consider the integral defining the polarization $P$ in Eq.~(\ref{eq:GWP})
first.
The Green function has poles (or cuts) just below the real $\omega$
axis for $\omega>\mu$ and above for $\omega<\mu$.
Therefore, if $z$ is a pole or a point in the cut, $\textrm{sgn}(\mu
-{\Re}z)=\textrm{sgn} ({\Im}z)$.
Note that the factor $e^{i\omega'\eta}$ implies that only the residues
related to occupied states ($\omega<\mu$) are summed up.
To avoid the just mentioned numerical difficulty, one can define the analytic
continuation of $P$ to complex frequencies as the sum over the same
residues, now evaluated at a complex frequency.\cite{gss88}
It is easy to show that for purely imaginary frequencies
this corresponds to rotate the integration contour to the complex
frequency axis $\mu+iu'$ ($u,u'$ real).
In the GWA the continued polarization [Eq.~(\ref{eq:GWP})] is:
\begin{eqnarray}\label{eq:m_con_p}
    P_0(iu) = -2i\int_{\mu-i\infty}^{\mu+i\infty} \mathrm{d}\omega'
    G_0(\omega'+iu)G_0(\omega').
\end{eqnarray}

On the same footing, also the self-energy [Eq.~(\ref{eq:GWS})] can be
continued to complex frequencies.
If $\omega=\mu+iu$, the following relation holds:
\begin{eqnarray}\label{eq:m_con_s}
    \Sigma_{\textrm{XC}}(\mu+iu) = i\int_{-i\infty}^{i\infty}
    \mathrm{d}\omega' G_0(\omega'+\mu+iu)W_0(\omega').
\end{eqnarray}
where the analytic continuation of $W_0(\omega'=iu')$ is evaluated by
inserting $P_0(iu)$ into Dyson's equation.
Note that the Lehmann representation of $P_0$ implies that a pole $z$
of $W_0$ has $\textrm{sgn}(\Im{z})=-\textrm{sgn}(\Re{z})$.

The self-energy resulting from Eq.~(\ref{eq:m_con_s}) will be known on
the complex line $\omega=\mu+iu$.
This is useful for the evaluation of integral properties (e.g., the
total energy), but for spectral properties the Green function (and
hence the self-energy) has to be known at real frequencies.
To this end, one can fit $\Sigma_{\textrm{XC}}$ on the complex axis
with a simple analytic expression, to be continued to the real
axis.\cite{rie99}
The multi-pole expansion is perhaps the more common one:
\begin{eqnarray}
    \Sigma_{\textrm{XC}}(\omega)=a_0+\sum_{j=1}^{N}\frac{b_j}{\omega-c_j}.
\end{eqnarray}
A small number of poles ($N=2\sim4$) normally provides a good fit.

Finally we recall a useful result to rotate the integration path in
frequency space.
Consider the two integrals, where $\omega$, $a$ and $b$ are real:
\begin{eqnarray}
    F_1(\omega)&=&\int_{-\infty}^{+\infty} d\omega'
    \frac{1}{(\omega'-z_1)(\omega+\omega'-z_2)}\nonumber\\ &=& i\pi
    \frac{{\textrm{sgn}}({\Im}z_1)-{\textrm{sgn}}({\Im}z_2)}
    {\omega-z_2+z_1}
\end{eqnarray}
\begin{eqnarray}
    F_2(\omega)&=&\int_{a-i\infty}^{a+i\infty}d\omega'
    \frac{1}{(\omega'-z_1)(b+i\omega+\omega'-z_2)}\nonumber\\ &=& i\pi
    \frac{{\textrm{sgn}}(a-{\Re}z_1)-{\textrm{sgn}}(a+b-{\Re}z_2)}
    {b+i\omega-z_2+z_1}
\end{eqnarray}
The two numerators are equal if
${\textrm{sgn}}({\Im}z_1)={\textrm{sgn}}(a-{\Re}z_1)$ and
${\textrm{sgn}}({\Im}z_2)={\textrm{sgn}}(a+b-{\Re}z_2)$.
In this case: $F_2(\omega)=F_1(b+i\omega)$, i.e., $F_2$ is the
analytic continuation of $F_1$ to complex frequencies $b+i\omega$.
Notice that, to be analytic, the continuation has to be performed {\em
after} the integration.

If both $z_1$ and $z_2$ are poles of the time-ordered Green function
[as for the polarization, Eq.~(\ref{eq:GWP})], it
follows from the Lehmann representation that the condition is met for
$a=\mu$ and $b=0$.
If $z_1$ is a pole of the effective interaction and $z_2$ is a pole of
the time-ordered Green function [as for the self-energy, Eq.~(\ref{eq:GWS})],
the condition is met for $a=0$, $b=\mu$.

\bibliography{gw_infinite}

\end{document}